\documentclass[12pt]{article}
\usepackage{amssymb}
\usepackage{amsfonts}
\begin{document}
\abovedisplayshortskip 12pt
\belowdisplayshortskip 12pt
\abovedisplayskip 12pt
\belowdisplayskip 12pt

\title{{\bf Supersymmetric and Deformed Harry Dym Hierarchies}}
\author{J. C. Brunelli$^{a}$, Ashok Das$^{b}$ and Ziemowit Popowicz$^{c}$  \\
\\
$^{a}$ Departamento de F\'\i sica, CFM\\
Universidade Federal de Santa Catarina\\
Campus Universit\'{a}rio, Trindade, C.P. 476\\
CEP 88040-900\\
Florian\'{o}polis, SC\\
Brazil\\
\\
$^{b}$ Department of Physics and Astronomy\\
University of Rochester\\
Rochester, NY 14627-0171\\
USA\\
\\
$^{c}$ Institute of Theoretical Physics\\
University of Wroc\l aw\\
pl. M. Borna 9, 50-205, Wroc\l aw\\
Poland}
\date{}
\maketitle

\begin{center}
{ \bf Abstract}
\end{center}

In this talk, we describe our recent results \cite{brunelli1,brunelli2}
on the supersymmetric
Harry Dym hierarchy as well as a newly constructed deformed Harry Dym
hierarchy which is integrable with two arbitrary parameters. In
various limits of these parameters, the deformed hierarchy reduces to
various known integrable systems.

\newpage

\section{Introduction:}

Supersymmetric extensions of a number of well know bosonic integrable
models have been studied extensively
in the past. The supersymmetric Korteweg-de Vries (sKdV)
equation~\cite{skdv}, the
supersymmetric nonlinear Schr\"odinger (sNLS) equation~\cite{snls} and
the supersymmetric Two-Boson
(sTB) equation~\cite{stb} represent just a few in this category. A
simple supersymmetric covariantization of
bosonic integrable models, conventionally known as the B supersymmetrization
(susy-B), has also attracted a lot
of interest because of the appearance of such models in string
theories. We  have, for instance, the B extensions of the KdV
(sKdV-B) equation~\cite{skdvb},
the supersymmetric TB (sTB-B) equation~\cite{stbb} and so on. Supersymmetric
extensions of integrable
models using a number $N$ of Grassmann variables greater than
one~\cite{n2skdv} and supersymmetric
construction of dispersionless integrable models~\cite{dispersionless}
have also been studied extensively in the past few years. The extended
supersymmetric models are particularly interesting because, in the
bosonic limit, they yield new classical integrable systems.

A classic bosonic integrable equation, the so called Harry Dym (HD)
equation~\cite{kruskal}, has
attracted much interest recently. The proprieties of
this equation are discussed in detail in
Ref. \cite{hereman}, and we simply emphasize that this equation shares the
properties typical of solitonic
equations, namely, it can be solved by the inverse scattering transform,
it has a bi-Hamiltonian
structure and infinitely many symmetries. In fact, the HD equation is
one of the most exotic
solitonic equations and the hierarchy to which it belongs, has a very
rich structure~\cite{brunelli}.
In this hierarchy we also have nonlocal integrable equations such as
the Hunter-Zheng (HZ)
equation~\cite{hunter}, which arises in the study of massive nematic
liquid  crystals as well as in
the study of shallow water waves. The HD equation, on the other hand,
is  relevant in the study of the Saffman-Taylor
problem which describes the motion of a two-dimensional interface
between a viscous and a non-viscous fluid~\cite{kadanoff}.

In
this talk we will describe systematically the question of
supersymmetrization  of the HD hierarchy as well as a deformation of
this hierarchy by two parameters that still is integrable. The talk is
organized as
follows. We will first review some of the essential features of the HD
equation and its hierarchy. Then, we will briefly describe the new
deformed model that is integrable. The simpler susy-B extension (sHD-B) of
the HD hierarchy as well its bi-Hamiltonian formulation and
Lax pairs will be discussed next. We will then derive the $N\!\!=\!\!1$
supersymmetric extensions of the HD (sHD) equation. We will show that, in
this case, there exist two nontrivial $N\!\!=\!\!1$
supersymmetrizations. In the
case of one of them, we have a bi-Hamiltonian description (we have not
found a Lax representation yet) while in the second case, we have a
Lax description (we have not found a Hamiltonian structure yet that
satisfies the Jacobi identity). We also describe the supersymmetric
extension for the HZ equation. Continuing on, we will describe the
$N\!\!=\!\!2$ supersymmetrization of the HD hierarchy which yields
four possibilities and we discuss their properties. 

\section{The Harry Dym Hierarchy:}

The Harry Dym equation
\begin{equation}
w_t=(w^{-1/2})_{xxx}\label{harrydym}\;,
\end{equation}
appears in many disguised forms, namely,
\begin{eqnarray}
v_t &=&{1\over 4}v^3 v_{xxx}\;,\nonumber\\
u_t&=&{1\over4}u^{3/2}u_{xxx}-{3\over 8}u^{1/2}u_xu_{xx}+{3\over
  16}u^{-1/2}u_x^3\label{liuhd}\;,\\
r_t&=&(r_{xx}^{-1/2})_x \;, \nonumber
\end{eqnarray}
where $v=-2^{1/3}w^{-1/2}$, $u=v^2$ and $r_{xx}=w$, respectively. In
this paper,
as in \cite{brunelli}, we will confine ourselves, as much as is
possible, to the form of the HD equation as given in (\ref{harrydym}).

The HD equation is a member of the bi-Hamiltonian
hierarchy of equations given by
\begin{equation}
w_t^{(n+1)}=\mathcal{D}_1 {\delta H_{n+1}\over\delta
w}=\mathcal{D}_2 {\delta H_{n}\over\delta w}\;,  \label{bihamiltonian}
\end{equation}
for $n=-2$, where the bi-Hamiltonian structures are
\begin{equation}
\mathcal{D}_1 = \partial^3,\qquad 
\mathcal{D}_2 = w\partial+\partial w\;,
\label{harrydymstructures}
\end{equation}
and the Hamiltonians for the HD equation are
\begin{equation}
H_{-1} =
\int  dx\,\left(2w^{1/2}\right),\quad 
H_{-2} = \int  dx\left({1\over8}w^{-5/2}w_x^2\right)\label{hdcharges}\;.
\end{equation}
We note here that the second structure in (\ref{harrydymstructures})
corresponds to the centerless Virasoro algebra while
\begin{equation}
{\cal D} = {\cal D}_{2} + c {\cal D}_{1}
\end{equation}
represents the Virasoro algebra with a central charge $c$. We note
also that the recursion operator following from
(\ref{harrydymstructures}),
$R=\mathcal{D}_2\mathcal{D}_1^{-1}$, can be explicitly inverted to yield
\begin{equation}
R^{-1}={1\over 2}\,\partial^3 w^{-1/2} \partial^{-1}w^{-1/2}\;.
\end{equation}
Furthermore, the conserved charges
\begin{eqnarray}
H_{0} &=&- \int  dx\,w \nonumber\;,\\
H_0^{(1)} &=&
\int  dx\,\left(\partial^{-1}w\right)\label{casimir}\;,\\
H_0^{(2)} &=& \int  dx\,\left(\partial^{-2}w\right)\;,\nonumber
\end{eqnarray}
are Casimirs (or distinguished functionals) of the Hamiltonian
operator ${\cal D}_1$ (namely, they are annihilated by the Hamiltonian
structure ${\cal D}_{1}$). As a consequence of this, it is possible to
obtain, in an explicit form, equations from
(\ref{bihamiltonian}) for integers $n$ both positive and negative, i.e.,
$n\in\mathbb{Z}$. As shown in
\cite{brunelli}, for $n>0$, we have three classes of nonlocal
equations. However, in this paper
we will only study the hierarchy associated with the local Casimir
$H_0$ in (\ref{casimir}). In
this way, for $n=1$, we obtain from (\ref{bihamiltonian}), with the
conserved charges
\begin{equation}
H_{1} = \int  dx\,{1\over
2}(\partial^{-1}w)^2,\quad 
H_{2} = \int  dx\,{1\over
2}(\partial^{-2}w)(\partial^{-1}w)^2\;,\label{nonlocalh}
\end{equation}
the Hunter-Zheng (HZ) equation
\begin{equation}
w_t=-(\partial^{-2}w)w_x-2(\partial^{-1}w)w\;,\label{hunterzheng}
\end{equation}
which is also an important equation that belongs to the Harry Dym hierarchy.

The integrability of the HD equation (\ref{harrydym}) also follows
from its nonstandard Lax representation
\begin{equation}
L = {1\over w}\partial^2,\qquad 
{\partial L\over\partial
t} = -2[B,L]\;,\label{lax}
\end{equation}
where
\begin{equation}
B=\left(L^{3/2}\right)_{\ge2}=w^{-3/2}\partial^3-{3\over
4}w^{-5/2}w_x\partial^2\;.\label{hdlaxb}
\end{equation}
Conserved charges, for $n=1,2,3,\dots$, are obtained from
\begin{equation}
H_{-(n+1)}=\hbox{Tr} L^{2n-1\over 2}\;.\label{negativehs}
\end{equation}
A Lax representation for the HZ equation (\ref{hunterzheng}) is also
known and is given by (\ref{lax}) with
\begin{equation}
B={1\over
  4}(\partial^{-2}w)\partial+{1\over4}\partial^{-1}(\partial^{-2}w)
\partial^2\;.\nonumber\\
\end{equation}
However, in this case, the operator $B$ is not directly related
to $L$, and, consequently, the Lax equation is not of much direct use
(in the construction of conserved charges etc).

\section{Deformed Harry Dym Hierarchy:}

Before going on to describe the supersymmetrization of this model, we
discuss briefly a new integrable model that can be thought
of as a deformation of the Harry Dym hierarchy. Identifying
$w=u_{xx}$, we note that the deformed Harry Dym and the deformed
Hunter-Zheng equations take the forms
\begin{eqnarray}
u^{\rm dHD}_{t} & = & \sqrt{2} \left(\frac{1-\lambda
  u_{xx}}{\sqrt{2u_{xx}-\alpha-\lambda
  u_{xx}^{2}}}\right)_{x}\;,\nonumber\\
u^{\rm dHZ}_{xxt} & = & \alpha u_{x} - 2u_{x}u_{xx} - u u_{xxx} +
  \frac{\lambda}{6} \left(u_{x}^{3}\right)_{xx}\;.\label{dHD}
\end{eqnarray}
Here $\alpha,\lambda$ are two constant parameters that cannot be
simultaneously scaled away. When $\alpha=\lambda=0$, these equations
reduce respectively to the Harry Dym and the Hunter-Zheng
equations. For $\lambda=0,\alpha\neq 0$, the deformed Hunter-Zheng
equation corresponds to the system studied by Alber et al
\cite{alber}. For $\alpha=1,\lambda\neq 0$, this sytem was considered
by Manna et al \cite{manna} and argued to be integrable. We have shown
\cite{brunelli2} 
that the system of equations in (\ref{dHD}) are bi-Hamiltonian and are
integrable for
arbitrary values of $\alpha,\lambda$ and belong to the same hierarchy.

\section{The Susy-B Harry Dym (sHD-B) Equations:}

The most natural generalization of an equation to a supersymmetric one
is achieved simply by working in a superspace. We note, from the HD
equation (\ref{harrydym}), that by a simple dimensional
analysis, we can assign the following canonical dimensions to various
quantities
\begin{equation}
[x]=-1\;,\quad[t]=3\;,\quad\hbox{and}\quad[w]=4\;.\nonumber
\end{equation}
The $N\!\!=\!\!1$ supersymmetric equations are best described
in the superspace parameterized by the coordinates
$z=(x,\theta)$, where  $\theta$ represents the Grassmann
coordinate ($\theta^2=0$). In this space, we can define
\begin{equation}
D={\partial\ \over\partial\theta}+\theta{\partial\ \over\partial
  x},\qquad D^{2} = \partial, \qquad [\theta] = -
  \frac{1}{2}\label{supercovariant}\;,
\end{equation}
representing the supercovariant derivative. 
Let us introduce the fermionic superfield
\begin{equation}
W=\psi+\theta w,\qquad [W] = [\psi] = \frac{7}{2}\label{superfield}\;.
\end{equation}

A simple supersymmetrization of a bosonic system, conventionally known
as the B supersymmetric (susy-B) extension \cite{skdvb}, is
obtained by simply replacing the bosonic variable $w$, in the original
equation, by 
\begin{equation}
(DW)=w+\theta\psi'\label{dsuperfield}\;,
\end{equation}
where $W$ represents a fermionic superfield. This leads to a manifestly
supersymmetric equation and following this for the case of the
equation (\ref{harrydym}), we obtain the susy-B HD (sHD-B) equation
\begin{equation}
W_t=\partial^2D\Bigl((DW)^{-1/2}\Bigr)\label{shd-b}\;,
\end{equation}
where $W$ is the fermionic superfield (\ref{superfield}).

This system is bi-Hamiltonian with the even Hamiltonian operators
\begin{equation}
\mathcal{D}_1 = \partial^2,\quad 
\mathcal{D}_2 = D(DW)D^{-1}+D^{-1}(DW)D\;,
\end{equation}
and the odd Hamiltonians (which follow from (\ref{hdcharges}) under the
substitution $w\rightarrow (DW)$)
\begin{equation}
H_{-1} = \int  dz\,2(DW)^{1/2},\quad 
H_{-2} =  \int  dz\,{1\over8}(DW)^{-5/2}(DW_x)^2\;.
\end{equation}
The Casimirs of ${\cal D}_1$ can be easily identified with the ones
following from (\ref{casimir}).

The sHD-B equation (\ref{shd-b}) has two possible nonstandard Lax
representations. Let
\begin{equation}
L= (DW)^{-1} D^{4} + c W_{x} (DW)^{-2} D^{3}.
\end{equation}
Then, it can be easily checked that the nonstandard Lax equation
\begin{equation}
\frac{\partial L}{\partial t} = \left[(L^{3/2})_{\geq 3} , L\right],
\end{equation}
leads to the sHD-B equation (\ref{shd-b}) for $c=0,-1$. Here the
projection $()_{\geq 3}$ is defined with respect to the powers of the
supercovariant derivative $D$. In a similar manner, we can obtain the
SHD-BB equations \cite{brunelli1}.

\section{The Supersymmetric $N\!\!=\!\!1$ Harry Dym (sHD) and
  Hunter-Zheng (sHZ) Equations:}

In this section, we will discuss $N\!\!=\!\!1$
supersymmetrization of the system and correspondingly, it is
appropriate to work in the superspace defined in
(\ref{supercovariant})--(\ref{superfield}). 

There are two basic ways
one can study the $N=1$ supersymmetrization of the Harry Dym
equation. First, we note that it is  possible
to supersymmetrize the two Hamiltonian structures of the Harry Dym
equation in
(\ref{bihamiltonian}), which is easily seen from the fact that the
second Hamiltonian structure is the centerless Virasoro algebra. Thus,
the supersymmetrized Hamiltonian structures follow to be
\begin{equation}
\mathcal{D}_1 = D\partial^2,\qquad
\mathcal{D}_2 =  \frac{1}{2}\left[W\partial+2\partial
  W+(DW)D\right]\;,\label{susystructures}
\end{equation}
and they are compatible.  Next, we write the most general local
equation in superspace that is consistent with the dimensional
counting which leads to a one parameter family of equations. Requiring
this to be bi-Hamiltonian with respect to 
(\ref{susystructures}), namely, requiring
\begin{equation}
W_t=\mathcal{D}_1 {\delta H_{-1}\over\delta
W}=\mathcal{D}_2 {\delta H_{-2}\over\delta W}\;,\label{susybihamiltonian}
\end{equation}
determines the parameter to be $a = 6$.
The Hamiltonians in (\ref{susybihamiltonian}), in this case have the
forms
($dz=dx\,d\theta$ with $\int d\theta=0$ and $\int d\theta\,\theta=1$)
\begin{eqnarray}
H_{-1} &=&
\int  dz\,2W(DW)^{-1/2}\nonumber\;,\\
H_{-2} &=& \int
dz\,{1\over8}\left[W_x(DW_x)(DW)^{-5/2}-15WW_xW_{xx}(DW)^{-7/2}\right]\;,
\end{eqnarray}
and the $N\!\!=\!\!1$ sHD equation assumes the simple form
\begin{equation}
W_t=D\partial^2\left(2(DW)^{-1/2}-3WW_x(DW)^{-5/2}\right)\;.\label{shd}
\end{equation}

It is easy to check that the Hamiltonian $H_{-1}$ is a Casimir of
${\cal D}_2$  and the conserved charge
\begin{equation}
H_0=-\int dz\,W
\end{equation}
is a Casimir of ${\cal D}_1$. As a result, the hierarchy can be
extended to positive values of $n$ and this leads to the $N=1$ susy
Hunter-Zheng equation
\begin{equation}
W_t=\mathcal{D}_1 {\delta H_{2}\over\delta
W}=\mathcal{D}_2 {\delta H_{1}\over\delta W}
=-{3\over2}W(D^{-1}W)-W_x(D^{-3}W)-{1\over2}(DW)(D^{-2}W)\;,
\label{hzbihamiltonian}
\end{equation}
where
\begin{eqnarray}
H_{1} &=&
\int  dz\,{1\over4}(D^{-1}W)(D^{-2}W)\nonumber\;,\\
H_{2} &=& \int  dz\,{1\over2}(D^{-1}W)(D^{-2}W)(D^{-3}W)\;.
\end{eqnarray}
Both the sHD and the sHZ equations are bi-Hamiltonian systems and the
infinite set of commuting conserved charges can be constructed
recursively. As a result, they decribe supersymmetric integrable
systems.

The second approach to finding a nontrivial $N\!\!=\!\!1$
supersymmetrization of the HD equation is to start with the Lax
operator in (\ref{lax}) and generalize it to superspace. Let us start
with the most general Lax operator involving non-negative powers of $D$,
\begin{equation}
L=a_{0}^2D^4+\alpha_1D^3+a_1D^2+\alpha_2D+a_2\;,\label{ansatz}
\end{equation}
where Roman coefficients are bosonic and Greek ones are
fermionic. It is easy to verify that, in this case, there are only
three projections, $(L^{\frac{3}{2}})_{\geq 0,1,3}$ (with respect to
powers of $D$), that can lead to a consistent Lax
equation. Using this ansatz,  we have not yet been able to
obtain the sHD equation (\ref{shd-b}) using fractional powers of the
Lax  operator (\ref{ansatz}). The Lax pair for this system, therefore, remains
an open question.

On the other hand, when
\begin{equation}
a_{0} = (DW)^{-1},\quad \alpha_{1} = c W_{x} (DW)^{-2},\quad a_{1} =
a_{2} = 0  = \alpha_{2},
\end{equation}
where $c$ is an arbitrary parameter, the nonstandard Lax equation
\begin{equation}
{\partial L\over\partial t}=[(L^{3/2})_{\ge3},L]\;,
\end{equation}
yields consistent equations only for $c=0,-1,-\frac{1}{2}$. As we have
pointed out in the last section, for the values of the parameter,
$c=0,-1$, we have the sHD-B equation. The third choice of the
parameter, therefore, leads to a new nontrivial $N\!\!=\!\!1$
supersymmetrization of the HD equation. Namely, with
\begin{equation}
L = (DW)^{-1} D^{4} - \frac{1}{2} W_{x} (DW)^{-2} D^{3}\;,\label{second}
\end{equation}
the Lax equation
\begin{equation}
\frac{\partial L}{\partial t} = \left[(L^{3/2})_{\geq 3}, L\right]\;,
\end{equation}
leads to a second $N=1$ supersymmetrization of the HD equation of the
form
\begin{eqnarray}
W_{t} & = & \frac{1}{16}\left[8D^{5} ((DW)^{-1/2}) - 3D(W_{xx}W_{x}
  (DW)^{-5/2})\right.\nonumber\\
 &  & \quad\left. + \frac{3}{4} (DW_{x})^{2} W_{x} (DW)^{-7/2} -
  \frac{3}{4} D^{-1}\left((DW_{x})^{3}
  (DW)^{-7/2}\right)\right]\;.\label{secondequation}
\end{eqnarray}
This is manifestly a nonlocal susy generalization in the variable $W$
which, however, is a completely local equation in the variable $(DW)$.

Since this system of equations has a Lax description, it is integrable
and the conserved
charges can be calculated in a standard manner and the first few
charges take the forms
\begin{eqnarray}
H_{1} & = & \int dz\, W_{x} (DW_{x}) (DW)^{-5/2}\;,\\
H_{2} & = & \int dz\, W_{x}\left[16 (DW_{xxx})(DW)^{-7/2} - 84
  (DW_{xx})(DW_{x})(DW)^{-9/2}\right.\nonumber\\
 &  & \left.\qquad + 77
  (DW_{x})^{3}(DW)^{-11/2}\right]\;,\nonumber
\end{eqnarray}
and so on. However, we have not yet succeeded in finding a
Hamiltonian structure
which satisfies Jacobi identity (it is clear that the Hamiltonian
structure is nonlocal, since the Hamiltonian is local).

\section{The $N\!\!=\!\!2$ Supersymmetric Harry Dym Hierarchy:}

The most natural way to discuss the $N=2$ supersymmetric
extension of the HD equation is in the $N=2$ superspace.
Just as we defined  a
superspace in the case of $N=1$ supersymmetry, let us define a
superspace parameterized by $z = (x, \theta_{1},\theta_{2})$,
where $\theta_{1},\theta_{2}$ define two Grasmann coordinates
(anti-comuting and nilpotent, namely,
$\theta_{1}\theta_{2}=-\theta_{2}\theta_{1}$, $\theta_{1}^{2} =
\theta_{2}^{2}=0$). In this case, we can define two supercovariant
derivatives
\begin{equation}
D_{1} =  \frac{\partial}{\partial \theta_{1}} + \theta_{1}
  \frac{\partial}{\partial x},\quad
D_{2} =  \frac{\partial}{\partial \theta_{2}} + \theta_{2}
  \frac{\partial}{\partial x}\;,\label{n2der}
\end{equation}
which satisfy
\begin{equation}
D_{1}^{2} = D_{2}^{2} = \partial\;,\quad D_{1}D_{2} + D_{2}D_{1} = 0.
\end{equation}
Such a superspace naturally defines a system with $N=2$
supersymmetry. Let us consider a bosonic superfield, $W$, in this
space which will have the expansion (we denote it by the same symbol
as in the case of $N=1$)
\begin{equation}
W=w_0+\theta_1\chi+\theta_2\psi+\theta_2\theta_1w_1\label{n2superfield}\;.
\end{equation}
Looking at the bosonic superfield $W$ in (\ref{n2superfield}), we
note that it has two bosonic components as well as two fermionic
components. In the bosonic limit, when we set the fermions to
zero, the $N=2$ equation would reduce to two bosonic equations.
Since we have only the single HD equation (\ref{harrydym}) to
start with, the construction of such a system is best carried out
in the Lax formalism. This also brings out the interest in such
extended supersymmetric systems, namely, they lead to new bosonic
integrable systems in the bosonic limit.

As in (\ref{ansatz}), let us consider the most general $N=2$ Lax
operator which contains differential operators in this
superspace of the following form (taking a more general Lax involving
only differential operators does not lead to equations which reduce to
the HD equation),
\begin{eqnarray}
L & = & W^{-1}\partial^{2} + (D_{1}W^{-1}) (\kappa_{1} D_{1} + \kappa_{2}
D_{2}) \partial + (D_{2}W^{-1}) (\kappa_{3} D_{1} + \kappa_{4} D_{2})
\partial\nonumber\\
 &  & \quad + \left(\kappa_{5} (D_{1}D_{2}W) W^{-2} + \kappa_{6}
   (D_{1}W)(D_{2}W) W^{-3}\right) D_{1} D_{2}\;,\label{n2lax}
\end{eqnarray}
where $\kappa_{i}, i=1,2,\cdots , 6$ are arbitrary constant
parameters. The $N=2$ supersymmetry corresponds to an internal $O(2)$
invariance that rotates $\theta_{1}\rightarrow \theta_{2},
\theta_{2}\rightarrow -\theta_{1}$ and correspondingly
$D_{1}\rightarrow D_{2}, D_{2}\rightarrow - D_{1}$ (thereby rotating
the  fermion components of the superfield
into each other). This invariance, imposed on the Lax operator,
identifies
\begin{equation}
\kappa_{4} = \kappa_{1},\quad \kappa_{3} = -\kappa_{2}\;.
\end{equation}

Using the computer algebra program REDUCE \cite{hearn} and the
special package SUSY2 \cite{susy2}, we are able to study
systematically the hierarchy of equations following from the Lax
equation
\begin{equation}
\frac{\partial L}{\partial t} = \left[(L^{3/2})_{\geq 2},
  L\right]\;.\label{n2laxequation}
\end{equation}

The consistency of the equation (\ref{n2laxequation}) leads to four
possible solutions for the values of the arbitrary parameters
\begin{enumerate}
\item $\kappa_{1} = \kappa_{2} = \kappa_{5} = \kappa_{6} = 0\;,$
\item $\kappa_{2} = 0$, $\kappa_{1} = \kappa_{5} = -
  \frac{\kappa_{6}}{2} = 1\;,$
\item $\kappa_{2} = \kappa_{5} = \kappa_{6} = 0$, $\kappa_{1} =
  \frac{1}{2}\;,$
\item $\kappa_{2} = 0$, $\kappa_{1} = \kappa_{5} = \frac{1}{2},
  \kappa_{6} = \frac{3}{4}\;.$
\end{enumerate}
The
first and the second cases can be checked to lead to the same
dynamical equation which is nothing other than the sHD-BB equations we
alluded to earlier.

For the third choice of parameters, we note that it leads to a
nontrivial $N=2$ supersymmetrization. However, 
in the bosonic sector, where we set all the fermions to zero, the
Harry Dym equation becomes decoupled from the other and, therefore, is
not very interesting. We also 
note that, under the $N=1$ reduction, it is straightforward to see
that the system goes over to the one discussed earlier.

The fourth case is probably the most interesting of all. Here, the Lax
operator takes the form
\begin{equation}
L^{(4)} =  - \left(W^{-1/2}D_{1}D_{2}\right)^{2}\;.\label{4lax}
\end{equation}
We note here that a similar situation also arises in the study of the
$N=2$ sKdV hierarchy \cite{n2susy} (for the case of the parameter
$a=4$). In this case, we have interesting nontrivial flows and, in the
bosonic limit, we have a new integrable system which has recently been
investigated further in \cite{sakovich}. We further note that
this supersymmetric system has a Hamiltonian structure of the form
\begin{equation}
{\cal D} = - 2 W^{\frac{1}{2}} D_{1}D_{2}\partial W^{\frac{1}{2}}\;.
\end{equation}

\section{Conclusions:}

In this talk, we have described our recent results
\cite{brunelli1,brunelli2} on the supersymmetrization as well as
deformation of the Harry Dym hierarchy.

\section*{Acknowledgments}

This
work  was supported in part by US DOE grant
no. DE-FG-02-91ER40685 as well as by NSF-INT-0089589.


\begin{thebibliography}{99}

\bibitem{brunelli1} J. C. Brunelli, A. Das and Z. Popowicz,
  {\em J. Math. Phys.\/} {\bf 44} (2003) 4756.

\bibitem{brunelli2} J. C. Brunelli, A. Das and Z. Popowicz, {\em
  Deformed Harry Dym and Hunter-Zheng Equations\/}, nlin.SI/0307043.

\bibitem{skdv} Y. I. Manin and A. O. Radul, {\em
  Commun. Math. Phys.\/} {\bf 98} (1985) 65;
P. Mathieu, {\em J. Math. Phys.\/} {\bf 29} (1988) 2499.

\bibitem{snls} G. H. M. Roelofs and P. H. M. Kersten, {\em
  J. Math. Phys.\/} {\bf 33} (1992) 2185;
J. C. Brunelli and A. Das, {\em J. Math. Phys.\/} {\bf 36}  (1995) 268.

\bibitem{stb} J. C. Brunelli and A. Das, {\em Phys. Lett.\/} {\bf
  337B} (1994) 303.

\bibitem{skdvb} K. Becker and M. Becker, {\em Mod. Phys. Lett.\/} {\bf
  A8} (1993) 1205.

\bibitem{stbb} J. C. Brunelli and A. Das, {\em Phys. Lett.\/} {\bf
  409B} (1997) 229.

\bibitem{n2skdv} C. Laberge and P. Mathieu, {\em Phys. Lett.\/} {\bf
  B215} (1988) 718; P. Labelle and
P. Mathieu, {\it J. Math. Phys.} {\bf 32} (1991) 923; Z. Popowicz,
  {\em Phys. Lett.\/} {\bf 174A} (1993) 411;
S. Krivonos, A. Pashnev and Z. Popowicz, {\em Mod. Phys. Lett.\/} {\bf
  A13} (1998) 1435.

\bibitem{dispersionless} J. Barcelos-Neto, A. Constandache and A. Das,
  {\em Phys. Lett.\/} {\bf A268}
(2000) 342; A. Das and Z. Popowicz, {\em Phys. Lett.\/} {\bf A272} (2000) 65;
A. Das and Z. Popowicz, {\em Phys. Lett.\/} {\bf A296} (2002) 15.

\bibitem{kruskal} M. D. Kruskal, Lecture Notes in Physics, vol. 38
  (Springer, Berlin, 1975) p. 310;
P. C. Sabatier, {\em Lett. Nuovo Cimento Soc. Ital. Fis.\/} {\bf 26}
  (1979) 477; {\it ibid} {\bf 26} (1979) 483;
L. Yi-Shen, {\em Lett. Nuovo Cimento Soc. Ital. Fis.\/} {\bf 70} (1982) 1.

\bibitem{hereman} W. Hereman, P. P. Banerjee and M. R. Chatterjee,
  {\em J. Phys.\/} {\bf A22} (1989) 241.

\bibitem{brunelli}   J. C. Brunelli and G. A. T. F. da Costa, {\em
  J. Math. Phys.} {\bf 43} (2002) 6116.

\bibitem{hunter} J. K. Hunter and Y. Zheng, {\em Physica\/} {\bf D79}
  (1994) 361.

\bibitem{kadanoff} L. P. Kadanoff, {\em Phys. Rev. Lett.\/} {\bf 65}
  (1990) 2986.

\bibitem{alber} M. S. Alber, R. Camassa, D. Holm and J. E. Marsden,
  {\em Proc. R. Soc. Lond.\/} {\bf 450A} (1995) 677.

\bibitem{manna} M. A. Manna and A. Neveu, {\em A singular integrable
  equation from short capillary-gravity waves}, physics/0303085.

\bibitem{hearn} A. Hearn, Reduce Manual Version 3.7, Santa Monica, CA
  (1999).

\bibitem{susy2} Z. Popowicz, Computer Physics Comm. {\bf 100} (1997)
  277.

\bibitem{n2susy} W. Oevel and Z. Popowicz, Comm. Math. Phys. {\bf 139}
  (1991) 441.

\bibitem{sakovich} S. Y. Sakovich, nlin.SI/0310039.



\end{thebibliography}
\end{document}